\documentclass[aps,twocolumn,prl,superscriptaddress,showpacs]{revtex4}
\usepackage[dvips]{graphicx}
\usepackage{amssymb,amsfonts,amsmath}

\usepackage{graphicx}

\begin{document}

\long\def\symbolfootnote[#1]#2{\begingroup%
\def\thefootnote{\fnsymbol{footnote}}\footnote[#1]{#2}\endgroup}

\newcommand{\avg}[1]{\langle{#1}\rangle}
\newcommand{\Avg}[1]{\left\langle{#1}\right\rangle}

\frenchspacing
\def\be{\begin{equation}}
\def\ee{\end{equation}}
\def\bc{\begin{center}}
\def\ec{\end{center}}
\def\bea{\begin{eqnarray}}
\def\eea{\end{eqnarray}}

\title{Assessing the relevance of node features for network structure}

\author{Ginestra Bianconi$^1$,
Paolo Pin $^{2,3}$ and Matteo Marsili$^1$}
\affiliation{$^1${The Abdus Salam International Center for Theoretical Physics, Strada Costiera 11, 34014 Trieste, Italy}\\
$^2${Dipartimento di Economia Politica, Universit\'a degli Studi di Siena, Piazza San Francesco 7, 53100 Siena, Italy}\\
$^3${Max Weber Programme, European University Institute, Via Delle Fontanelle 10, 50014 San Domenico di Fiesole (FI), Italy}}

%\footlineauthor{Bianconi, Pin and Marsili}

%\contributor{Submitted to Proceedings of the National Academy of Sciences
%of the United States of America}

%\conflictofinterest{Author contributions: G.B., M.M. and P.P. designed research, performed research, wrote the
%paper and analyzed data.}

%\conflictofinterest{The authors declare no conflict of interest.}

%\maketitle

%{\footnotesize
%Classification: Social Sciences
%}
%\vspace{10pt}

\begin{abstract}
Networks describe a variety of interacting complex systems in social
science, biology and information technology. Usually the nodes of
real networks are identified not only by their connections but also by some other
characteristics.
Examples
of characteristics of nodes  can be age, gender or nationality of a person in a
social
network, the abundance of  proteins in the cell taking part in a
protein-interaction networks or the geographical position of airports
that are connected by directed flights.
Integrating the information on the connections of each node with the
information about its  characteristics is crucial to discriminating between the essential
 and negligible characteristics of nodes for the structure of the network.
In this paper  we propose a general indicator $\Theta$, based on entropy measures,
to quantify the dependence of a network's structure on a given set of  features.
 We apply this method to social networks of
friendships in US schools, to the
protein-interaction network of {\it Saccharomyces cerevisiae} and to the US airport
network, showing
that the proposed measure provides information which complements other known measures.
\end{abstract}

%\keywords{ networks | entropy | inference  }
\maketitle
%\abbreviations{\dots}

%\body

{N}etworks have become a general tool for describing the structure of
interaction or dependencies in such disparate systems as cell
metabolism, the internet and society \cite{RMP,Doro,Newman_rev,Latora,
  Caldarelli}. Loosely speaking, the topology of a given network can
be thought of as the byproduct of {\em chance and
  necessity} \cite{Monod}, where functional aspects and structural
features are selected in a stochastic  evolutionary process. The issue of separating
 ``chance'' from ``necessity'' in networks has attracted much interest.
This entails understanding random network ensembles (i.e. chance) and
their inherent  structural features \cite{loopsGBMM,cliquesGBMM,Reichardt} but also developing techniques to infer
structural and functional characteristics on the basis of a given network's topology.
Examples go from inference of gene function from protein-interaction networks
\cite{Leone} to the detection of communities in social networks
\cite{Fortunato_rev,Danon}.
Community\footnote{A community structure, in general terms, is an assignment of nodes into classes. Community detection aims at partitioning nodes into homogeneous classes, according to similarity or proximity considerations.} detection, for example, aims at uncovering a hidden classification of nodes, and
a variety of methods have been proposed relying on {\em i)} structural properties of the network
(betweenness centrality \cite{GN}, modularity \cite{Modularity}, spectral
decomposition \cite{spectral}, cliques \cite{Vicsek} and hierarchical structure
\cite{Newman_hierarchical}), {\em ii)} statistical methods \cite{Newman}
or {\em iii)} on processes defined on the network \cite{Arenas,Reichardt}.
Implicitly, each of this method relies on a slightly different understanding of what a community is.
Furthermore, there are
intrinsic limits to detection; often  the outcome  depends on the algorithm and a clear
assessment of the role of chance is possible in only a few cases (see e.g.
\cite{Reichardt,Fortunato}).

%For the case of community detection, in particular, a
%variety of methods have been proposed relying on {\em i)} structural properties such
%as betweenness centrality \cite{GN}, modularity \cite{Modularity}, spectral
%decomposition \cite{spectral}, cliques \cite{Vicsek} and hierarchical
%\cite{Newman_hierarchical} structure, {\em ii)} statistical methods \cite{Newman}
%or {\em iii)} on processes defined on the network \cite{Arenas,Reichardt}.
%Implicitly, all these methods attempt to uncover a hidden classification of nodes
%which might be of relevance to the specific system under study. But there are
%intrinsic limits to detection; often  the outcome  depends on the algorithm and a clear
%assessment of the role of chance is possible in only a few cases (see e.g.
%\cite{Reichardt,Fortunato}).

As a matter of fact, in several cases a great deal of additional
information, beyond the network topology, is known about the
nodes. This comes in the form of attributes such as age, gender and
ethnic background  in social networks, or annotations of known functions for
genes and proteins. Sometimes this information is incomplete, so
it is legitimate to attempt to estimate missing information from the
network's structure. But often the empirical data on the network are no more
reliable or complete than those on the attributes of the nodes.
In such cases, it may be more informative to ask what the functions or attributes of the nodes tell us about the network than the other way around.
In this paper we propose an indicator $\Theta$ that quantifies how
much the topology of a network depends on a particular assignment of
node characteristics. This provides an information bound which can be used as a benchmark for feature extraction algorithms.
This exercise, as we shall see, can also reveal statistical regularities which shed light on possible mechanisms underlying network's stability and formation.

In the following, we first define  $\Theta$, then we investigate
separately the
case in which node characteristic assignment induces a community structure on the
network, and the case in which the assignment corresponds to a
position of the nodes in some metric space.
We will calculate $\Theta$ for benchmarks and for examples of  social,
biological and economics  networks.

\section{Definition of $\Theta$}

We shall first give a description of our indicator $\Theta$ in a simple
case study and then give a
general abstract definition.

Let us  consider the specific problem of evaluating the significance of the
network community structure $\vec{q}=(q_1,\ldots,q_N)$ induced  by the assignment of a characteristic $q_i\in\{1,\ldots,Q\}$, to each node $i\in\{1,\ldots,N\}$ of a network of $N$ nodes.
Individual nodes are characterized by their degree $k_i$, which is the number of links they have to other nodes in the network.
The network $g$ is fully specified by the adjacency matrix taking values $g_{i,j}=1$ if nodes $i$ and $j$ are linked and $0$ otherwise.
The community structure induced by the assignment $q_i$ on the network
is described by a matrix $A$ of elements
$A(q,q')$ indicating the total number of links between nodes with
characteristics $q$ and $q'$.
A natural measure of the significance of the induced community structure
$\vec q$ on the network $g$ is provided by the number of graphs $g'$ between those individual nodes (characterized by the degree sequence $\vec k$) which are consistent with $A$.
The logarithm of this number is the entropy $\Sigma_{\vec k,\vec q}$ \cite{entropy,entropy2} of the distribution which assigns equal weight to each graph $g$ with the same $\vec q$ and $\vec k$.
This number also depends on the degree sequence $\vec k$ and the relative frequency of different values of $q$ across the population.
These systematic effects are removed considering the entropy
$\Sigma_{\vec k,{\pi}(\vec{q})}$ obtained from a random permutation $\pi(\vec{q}): i\to q_{\pi(i)}$
of the assignments, where $\{\pi(i), i=1,\ldots,N\}$ is a random permutation of the integers $i\in\{1,\ldots,N\}$.
The indicator $\Theta$ is obtained as the standardized deviation of  $\Sigma_{\vec k,\vec q}$ from the entropy $\Sigma_{\vec k,{\pi}(\vec{q})}$ of networks with randomized assignments:
\begin{equation}
\Theta_{\vec{k},\vec{q}}=\frac{E_\pi\left[\Sigma_{\vec{k},\pi(\vec{q})}\right]-\Sigma_{\vec{k},\vec{q}}}
{\sqrt{
E_\pi\left[\left(\Sigma_{\vec{k},\pi(\vec{q})}-E_\pi\left[\Sigma_{\vec{k},\pi(\vec{q})}\right]\right)^2\right]
}} \ ,
\label{ThetaA}
%\label{theta}
\end{equation}
where $E_\pi[\ldots]$ stands for the expected value of over random uniform permutations $\pi(\vec q)$ of the assignments.
In words, $\Theta$ measures the specificity of the network $g$ for the particular assignment $\vec q$, with respect to assignments obtained by a random permutation.

The indicator $\Theta$ can be similarly defined in a much more general setting, with the following  abstract definition:
 Let $g\in {\cal G}_N$ be the network we are
interested in, where $N$ is the number of vertices and $g_{i,j}$ is the
adjacency matrix.
${\cal G}_N$ is the set of all graphs of $N$ vertices.
An assignment is a vector $\vec{q}$, such that for each node $i$, $q_i\in Q$ is
defined on a set $Q$ of possible characteristics, given by the context.
Call ${\cal Q}=Q^{N}$ the set of all possible such vectors on $Q$.
A feature is a mapping $\phi: {\cal G}_N\times{\cal Q}\to\Phi$, which associates to
each graph $g$ and assignment $\vec{q}$ a graph feature $\phi(g,\vec{q})\in \Phi$.
As will become clear, we do not need any assumption about the topology of the set of
features $\Phi$.

A simple example of  features is those which do not depend on any assignment
($\phi(g,q)=\phi(g)$), such as the number of edges, or the degree sequence.
Instead, the previously introduced  community structure $A$ is an example of a feature  depending  both on the degree sequence $\vec k$ and on the
assignment $\vec{q}$, i.e. $\phi(g,\vec{q})=\{\vec k,~A(q,q'),~q,q'\in Q\}$.\footnote{To be precise, here $k_i=\sum_j g_{i,j}$ is the degree and $A(q,q')=\sum_{i,j}g_{i,j}\delta_{q_i,q}\delta_{q_j,q'}$ is the number of links between nodes with attribute $q$ and $q'$.}

In order to assess the relevance of a feature  $\phi(g,\vec{q})$,   we make
use of the entropy
$\Sigma_{\phi(g,\vec{q})}$ of randomized network ensembles \cite{entropy,entropy2}.
The entropy of the ensemble of graphs with feature $\phi(g,\vec{q})$ is defined as
the normalized  logarithm of the number of possible graphs, consistent
with $\phi(g,\vec{q})$ and normalized by $N$:\footnote{In other words, $\Sigma_{\phi(g,\vec{q})}$ is the Gibbs-Boltzmann entropy of the ensemble of graphs which assigns  equal weight to each graph $g$ satisfying the constraints, which is equivalent to the usual Shannon entropy of the distribution of graphs in this ensemble.}
\begin{equation}
\label{Sigma}
\Sigma_{\phi(g,\vec{q})}=\frac{1}{N}\log \left|\{g'\in{\cal
G}_V:~\phi(g',\vec{q})=\phi(g,\vec{q})\}\right| \ .
\end{equation}
This quantity evaluates the level of randomness that is
present in the ensemble of networks with a given feature. The
numerical evaluation of the entropy  $\Sigma_{\phi(g,\vec{q})}$ is a
very challenging problem. On the contrary this quantity can be
theoretically  calculated by introducing a partition function in a
statistical mechanics formalism and  evaluating it  by saddle point
approximation (see the Supplementary Materials for the equations and
the codes for the evaluation of  $\Sigma$ ).
Finally, with the same notations used above, the indicator $\Theta$ is defined as
\begin{equation}
\label{Theta}
\Theta_{\phi(\vec{k},\vec{q})}=\frac{E_\pi\left[\Sigma_{\phi(\vec{k},\pi(\vec{q}))}\right]-\Sigma_{\phi(\vec{k},\vec{q})}}
{\sqrt{
E_\pi\left[\left(\Sigma_{\phi(\vec{k},\pi(\vec{q}))}-E_\pi\left[\Sigma_{\phi(\vec{k},\pi(\vec{q}))}\right]\right)^2\right]
}} \ .
%
%\Theta_{g,\vec{q}}=\frac{\Sigma_{\phi(g,\vec{q})}-\langle
%\Sigma_{\phi(g,\pi(\vec{q}))}\rangle_\pi}
%{\sqrt{\langle \delta\Sigma_{\phi(g,\pi(\vec{q}))}^2\rangle_\pi}} \ ,
%\label{theta}
\end{equation}
The quantity $\Theta$   provides a measure of the relevance of a given feature
$\phi(g,q)$ for the structure of the network.
While $\Sigma_{\phi(g,q)}$ can be obtained in analytic form, the average and the standard deviation over
permutations require a random sampling of the space of possible
permutations of the characteristics. In practice, $N_{\rm samp}$ random permutations are drawn in order to estimate the expected value and the variance of $\Sigma_{\phi(\vec{k},\pi(\vec{q}))}$ in Eq. (\ref{Theta}). Furthermore, the maximal deviation of $\Sigma_{\phi(\vec{k},\pi(\vec{q}))}$ from the expected value provides an estimate of the confidence interval at probability $p=1/N_{\rm samp}$.\footnote{A more precise estimate of the probability of occurrence of a given value of $\Theta$ would entail the study of large deviation properties of the entropy distribution. This goes beyond our present purposes.}

Besides the value of $\Theta$, our approach also provides more detailed information.
Technically, this is extracted from the saddle point values of the Lagrange multipliers introduced in the calculation of $\Sigma_{\phi(g,q)}$ in order to enforce the constraints (see Supplementary Material). In the examples discussed below, this information is encoded in the probability  that a node $i$ is linked to a node $j$ in an
ensemble with a given feature  $\phi(g,\vec{q})$. This is given by
\be
p_{ij}=\frac{z_i z_j W(q_i, q_j)}{1+z_i z_j W(q_i, q_j)} \ \ .
\label{community}
\ee
The value of the ``hidden variables'' $\vec{z}$ and the statistical
weight $W(q,q')$ can be inferred from the real data
\cite{entropy,entropy2}. Therefore the function $W(q_i,q_j)$ can shed light on the dependence of the probability of a link between nodes $i$ and $j$, on their assignments $q_i$ and $q_j$.

\begin{figure}[t]
\includegraphics[width=0.42\textwidth]{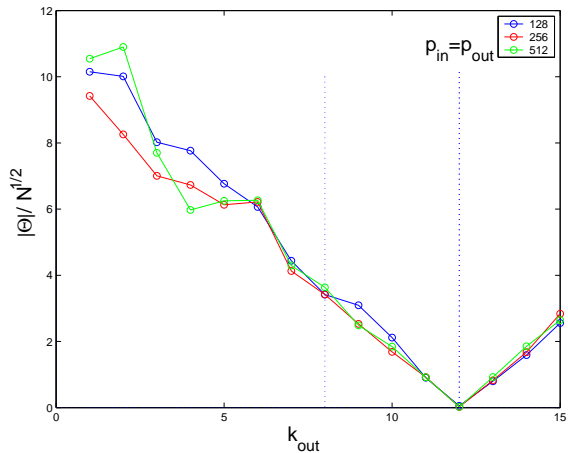}
\caption{The dependence of  $\Theta$  on $k_{out}$ in
  random networks of  $N=128,~256,~512$ nodes with four equal size communities and
  average connectivity $\bar{k}=16$ (each point is obtained as an averages over $10$ realizations).
  To be compared with Fig.7 in \cite{Modularity} and Fig.10 in \cite{Reichardt}.\label{thetak.fig}}
\end{figure}

\section{Application to networks with a community structure}

In the following, we  will describe how to measure $\Theta$ for
assessing the relevance of a community structure.
First, we analyze the behavior of $\Theta$ on synthetic data-sets. These have been used as benchmarks for community detection algorithms \cite{Modularity,Reichardt}. For these benchmarks we find that $\Theta$ increases with the number $N$ of vertices, reflecting the intuitive idea that larger graphs can resolve finer information on the global architecture of the network.
%
%Here these benchmarks will allow us to to quantify the dependence of $\Theta$ on the number $N$ of vertices.
We shall see that even in the region where community detection algorithms fail, there is a detectable influence of community structure on the topology of the network.
Next, we apply this tool to a social network and a biological
network. In particular, we will consider a dataset of  friendship networks
in US schools  and a network of high confidence protein-protein
interaction \cite{Proteins}.
The dataset of friendship networks in US schools, which includes $84$
schools, is particularly suitable for contrasting the information gained
from $\Theta$ to  that derived from other indicators, such as modularity \cite{Modularity}.
We will show that, at least in this case study,  the  information provided by $\Theta$ is of a different nature and more detailed than that provided by other measures.

As discussed above, in this section we shall take
$q_i\in\{1,\ldots,Q\}$ to be the label of the class which node $i$
belongs to, with $Q< \sqrt{N}$.\footnote{This  limitation is imposed by
the fact that the saddle point method we use to evaluate the entropy
is reliable only if the number of imposed constraints $N+Q^2$ is of the same
order of magnitude of  $N$.}
The feature
$\phi(g,\vec{q})=\{\vec{k}, A(q,q')\}$ specifies the degree sequence
$\vec k$ and the number $A(q,q')$ of links between nodes in
communities $q$ and  $q'$.
Finally, we calculate the indicator
$\Theta$ defined in Eq. (\ref{Theta})
for the different cases.

\subsection{Evaluation of $\Theta$ on benchmarks}

We evaluate $\Theta$ on the benchmark random networks, originally proposed in Refs. \cite{Modularity,Reichardt}, of $N=128,~256,~512$
nodes divided into four communities of equal size with fixed average
connectivity $\bar{k}=16$, varying  the average degree $\bar{k}_{out}$
towards different communities.

The results are shown in Fig. $\ref{thetak.fig}$. This shows that, for a fixed structure, $\Theta$ for different values of $N$ nicely collapses on a single curve when rescaled by the factor $\sqrt{N}$. This suggests that the size dependence of $\Theta$ results from the random fluctuations of the intensive quantity $ \Sigma_{\phi(g,\pi(\vec{q}))}$. Hence the same scaling is expected in general, in not too heterogeneous systems.\footnote{A plausibility argument for the scaling behavior is the following:
%Consider two random permutations $\pi_0$ and $\pi_N$ of the assignments, and the difference $\Sigma_{\pi_N}-\Sigma_{\pi_0}$ of the associated entropies. $\pi_N$ can be realized, starting from $\pi_0$ in many possible ways by successive exchanges of assignments on pairs of nodes. Each of these defines a path $\{\pi_n,~n=1,\ldots,N\}$ in the space of permutations and the entropy difference can be written as the sum of the entropy differences on the $N$ individual "steps". The expected value of $(\Sigma_{\pi_N}-\Sigma_{\pi_0})$ can then be expanded as a double sum of
%E[($\Sigma_{\pi_n}-\Sigma_{\pi_{n-1}})($\Sigma_{\pi_n'}-\Sigma_{\pi_{n'-1}})]$ for $1\le n,n'\le N$. The off diagonal terms entail exchanges in the assignments of different nodes, which will generally be far apart, and their effect will average to zero. The diagonal term, instead, yields a term of order $N/
Consider a particular permutation $\pi$ and imagine to make a small number $n\ll N$ of further perturbations, by exchanging assignments on pairs of randomly chosen nodes. Each such perturbation is likely to affect a different part of the network, which means that the associated changes in the entropy can be considered as uncorrelated. Hence we expect a change in the entropy density of the order of $\sqrt{n}/N$. This is expected to hold true also for $n/N$ finite but small suggesting that, as $N$ increases, the difference between the entropies of two random permutations -- and hence the denominator in Eq. (\ref{Theta}) -- is of order $1/\sqrt{N}$.}

Secondly, Fig. \ref{thetak.fig} shows that $\Theta$ vanishes only
when there is no distinction in linking probabilities: with 4 groups this occurs when $\bar{k}_{out} \simeq \frac{3}{4} \bar{k}=12$, which is larger than the value ($\bar{k}_{out}\approx 8$) where community detection algorithms fail \cite{Modularity,Reichardt}. Indeed, community detection can be seen as the inverse problem to that addressed here. In this spirit, $\Theta$ provides an {\em a-priori} bound on the possibility of detecting communities in networks, as well as a universal indicator of the performance of different algorithms.

\subsection{The dataset of friendship networks in US  schools}

We apply our method to a dataset of $84$ US schools in which
students were asked to provide information about themselves
(among other things specifying in particular sex,
age, and ethnic background) together with the  names
of up to 5 of their female friends and up to 5 of their male friends.
Although the networks are directed in origin, in our analysis, in
order to simplify study, we consider them as
undirected, where each undirected link is present if at least one of
the two students has indicated the linked one as his/her friend.
The maximal connectivity of these networks is $k=16$, reached only
in rare cases.
This dataset is particularly interesting for the study of
homophily in American schools \cite{Moody01,Kertesz,Pin}.

We have measured the value of $\Theta$ associated with the
community structure induced by    the self-reported
ethnic background  (there were six possibilities in the
questionnaire: $Q=6$)  in all the $84$ schools of the dataset.
Loosely speaking, in this case $\Theta$ measures the extent to which ethnic background shapes the  social network
of friendship in US schools.

For 25\% of the schools we find that $\Theta$ is not significative, at the 5\% confidence level.  For the rest of the data set, $\Theta$ takes widely scattered values across schools (up to $\Theta\simeq 532$).
In order to asses how much the variation in $\Theta$ correlates with  ethnic diversity we take, as a measure of diversity in the assignment $\vec{q}$, the Shannon entropy
\be
S=-\sum_{q=1}^Q x_q \log(x_q) \ \ ,
\ee
where $x_q$ is the fraction of nodes with  $q_i=q$. We remark that the Shannon entropy $S$ of a partition measures the diversity in the population but does not contain any information on the social network.

\begin{figure}[t]
%\vspace{5pt}
\includegraphics[width=0.42\textwidth]{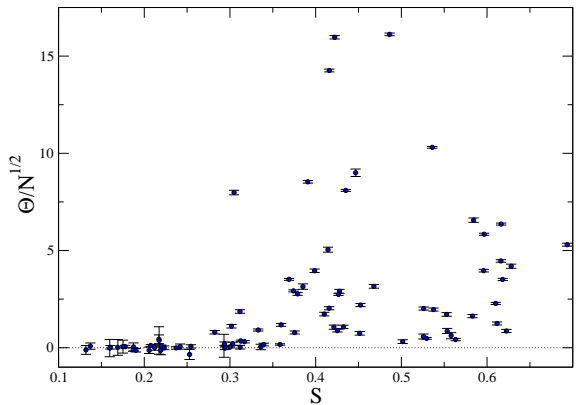}
%\centerline{\psfig{file=Theta_S.eps,width=9cm}}
\caption{Relation between  $\Theta/\sqrt{N}$, which captures the relevance of ethnic background for friendship networks, and  the Shannon entropy $S$, indicating the ethnic diversity of the student population. Each point corresponds to a different US schools in the data-set. Error bars indicate 5\% confidence intervals.\label{ThetaS}}
\end{figure}

In Fig. \ref{ThetaS} we report the dependence of $\Theta$ on
$S$. We observe that the value of $\Theta/\sqrt{N}$ is small and not statistically significant in ethnically uniform schools ($S<0.3$) but it grows larger and significant for schools with a stronger diversity. The largest values of $\Theta$, as well as the largest spread, occur for intermediate values of $S$ ($0.4 \div 0.5$), suggesting a nontrivial dependence.
In order to asses the statistical relevance of this result, we have studied the dependence of $\Theta$ on $S$ in benchmark synthetic networks, such as the ones presented above, where the fraction of links within the community of each individual is kept constant, but the relative sizes of communities are varied. A much weaker, barely significant increase of $\Theta$ with $S$ was found in synthetic networks, hinting that a non-trivial interplay between homophily and diversity might be responsible for the features observed in Fig. \ref{ThetaS}.

%in this dataset,
%more diversified is the student population (higher Shannon entropy $S$) more
%segregated are the friendhip communities (higher value of
%$\Theta/\sqrt{N}$).
%This  trend indeed follow  the law
% $\Theta/\sqrt{N}\sim N^{\delta}$ with
%$\delta=3.1 \pm 0.1$. Variations of $\Theta/\sqrt{N}$ from this trend
%indicates schools more or less segregated respect to the general
%trend.

A popular measure for community structure, frequently used in the literature, is modularity, which is closely related to inbreeding homophily indices in social sciences \cite{Coleman58} and the $F$-statistic in genetics \cite{Wright22}.

Modularity $M$  measures how densely connected the nodes that belong to the same partition are. It is  defined as
\be
M=\sum_{q=1}^Q\left[\frac{l_q}{L}-\left(\frac{k_q}{2L}\right)^2\right] \ \ ,
\ee
where $Q$ is the total number of communities or classes, $L$ is the
total number of links, $l_q$ is the
total number of links  joining nodes of community $q$ and $k_q$ is the
sum of the degrees of the nodes  in the community $q$.

\begin{figure}[t]
\includegraphics[width=0.42\textwidth]{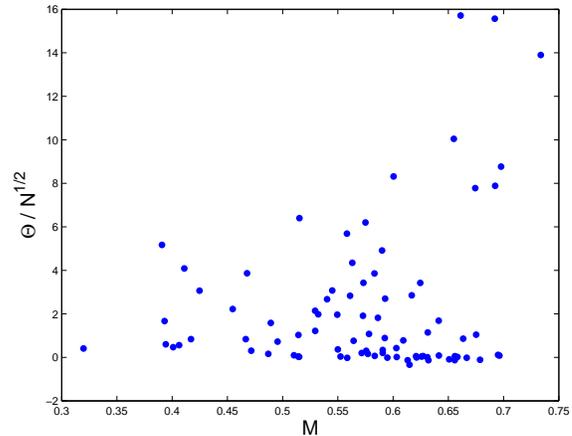}
%\centerline{\psfig{file=figure1.eps,width=7cm}}
\caption{The value of $\Theta/\sqrt{N}$ versus the modularity $M$ for the
dataset of friendship networks in American Schools. Each point is a school.\label{fig1}}
\end{figure}

Fig. $\ref{fig1}$ reports the value of $\Theta/\sqrt{N}$ versus the value of the modularity $M$ for each school, suggesting that the  two indicators are not simply correlated.  The two indicators provide different information: loosely speaking, while $M$ provides an absolute measure of the excess of inward or outward links in a community assignment, $\Theta$ measures how much the biases in the community assignment is correlated with the network topology.
%
%$\Theta$ provides a qualitatively different information on the relevance of community structure and, in particular, it is more sensitive to the topology of the network. Indeed, a community assignment with a given value of the modularity, is more informative on the network topology when the network is strongly clustered in groups, than when the network has a less pronounced cluster structure.

\begin{figure}[t]
\includegraphics[width=0.42\textwidth]{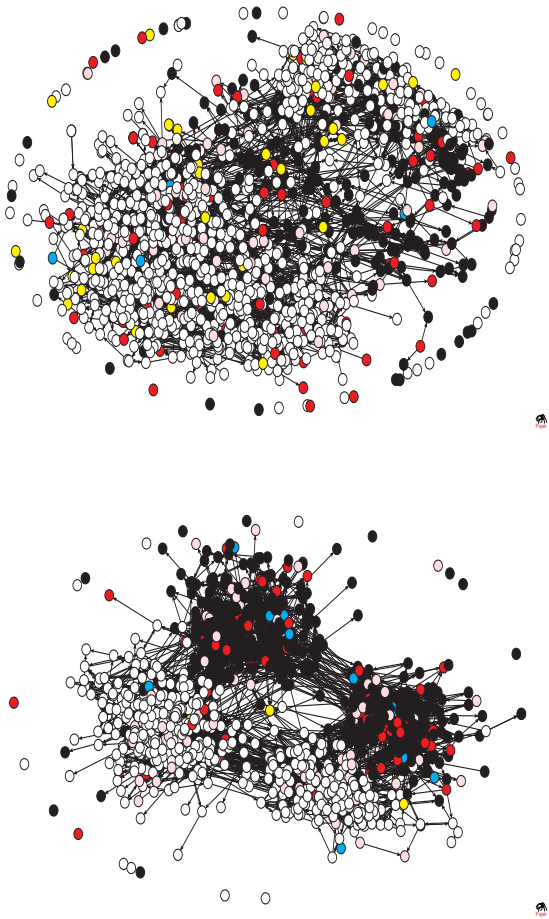}
%\centerline{\psfig{file=schools3.eps,width=7cm}}
\caption{ The case of two schools with similar modularity and Shannon
  entropy but very different value of $\Theta$. The top figure
  represents the friendship network of a school  of $N_1=1461$ students, average
connectivity $\avg{k}=5.3$,
  Shannon entropy $S_1=0.41$, modularity $M_1=0.64$ and
  $\Theta_1/\sqrt{N}=1.69$. The bottom figure represents the friendship network of
  a school of $N_2=1147$ students, average degree $\avg{k}=8.8$,
  Shannon entropy $S_2=0.48$, modularity $M_2=0.66$ and
  $\Theta_2/\sqrt{N}=15.71$. The different colors represent the
  self-reported ethnic backgrounds of the students.\label{fig2}}
\end{figure}

In order to substantiate this statement in a visual manner, we identify two schools with different values of $\Theta/\sqrt{N}$, but similar values of $N$, modularity $M$ and Shannon entropy $S$.
Fig. \ref{fig2} reports the friendship networks in the two schools,
strongly suggesting that significant differences in $\Theta$ imply different degrees of separation between the different communities, an effect which is not captured by $M$. This shows that a community assignment with a given value of the modularity, is more informative on the network topology when the network is strongly clustered in groups, than when the network has a less pronounced cluster structure.
%\bigskip

\subsection{The dataset of a protein-protein interaction network}

We apply the proposed method to the study of the relevance
of the protein abundance on the protein
interacting map of {\it Saccharomyces cerevisiae}.
The dataset, published in \cite{Proteins}, is  a subset of
the protein-interaction network of {\it Saccharomyces cerevisiae} formed
by $N=1,740$ proteins with known concentrations $x_i$ and $4,185$ interactions,
independently confirmed in at least two publications.
The abundance of a protein varies between 50 molecules per cell up to
1,000,000 molecules per cell with a median of 3,000 molecules per cell.
The  abundance of a protein is not correlated with simple local  structural features of the protein interaction map, such as the degree ($R=0.13$) or the clustering coefficient ($R=0.005$).
This raises the question of whether the concentration of proteins has any relevance to the interaction network and if so, what information it provides.

\begin{figure}[t]
\includegraphics[width=0.42\textwidth]{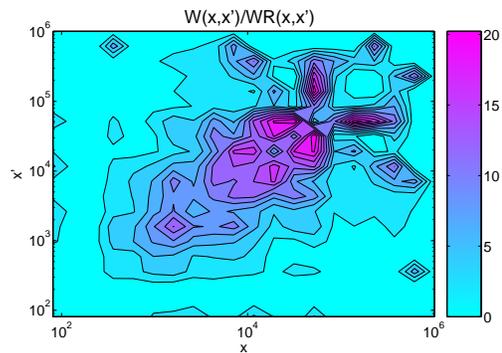}
%\centerline{\psfig{file=Wn.eps,width=9cm}}
\caption{Relevance of protein abundance $x$ for the protein-protein interaction
  network studied in \cite{Proteins}. The statistical weight $W(x,x')$ describing the likelihood of links between proteins with concentrations $x$ and $x'$ is first normalized to the analogous function $WR(x,x')$ which is obtained in the randomized data-set (with a random permutation of the abundance values $x_i$). The density plot reports the dependence of $W(x,x')/WR(x,x')$ as a function of the protein abundance $x$, $x'$.
%\vspace{15pt}
\label{Wxx'}}
\end{figure}

We bin the abundance $x$ into 20 logarithmically spaced
intervals given by the ordered vector $\vec{x}=(x_0,x_1,\ldots,x_{20})$.
Next, we assign to each protein $i$ the
corresponding coarse--grained abundance $q_i={k}$ if $x_i\in [x_{k-1},x_{k})$.
The features of the network that we consider are again the connectivity
of each protein together with  the number of links
between proteins of different abundance $A(q_i,q_j)$.
We find a value of $\Theta=21.76$, well beyond the 1\% confidence interval
$\Theta<2.7$, showing that the abundance of
the protein encodes relevant information on the network structure.
In Fig. $\ref{Wxx'}$ we report the  value of the statistical weight $W(x,x')$ in Eq. (\ref{community}) as a function of the (log--) abundance
%$x=\log c$ and $x'=\log c'$
of each pair of proteins in the network. The value of  $W(x,x')$ is normalized to the  value $WR(x,x')$  found in networks where the protein abundance is randomized, in order to highlight features of the specific concentration assignments in the data-set. The maximum of $W(x,x')/WR(x,x')$ along the diagonal suggests that proteins of a given concentration tend to interact preferentially with proteins with a similar concentration, therefore showing some ``assortativity'' of the
interaction map in the plane of the abundance $x,x'$.

\section{Application to spatial networks}

The role of the space in which networks are embedded, and its implications on navigability
and efficiency, has attracted considerable interest \cite{Yook,Kleinberg,Boguna,Delos}.
Here we show how the proposed indicator $\Theta$ can be used for assessing how
relevant the spatial position of the nodes in some geographical or
abstract metric space is.

In this case, each node can be  characterized by its
degree $k_i$ and by its position in space $q_i$. We first define a
set    $d\in\{d_1,\ldots, d_D\}$ ($D={\cal O}(N)$) of  fixed increasing
distance values. We then consider the ensemble of networks with given
feature   $\phi(g,\vec{q})=\{ \vec{k}, B(d) \}$, where
$B(d)=(b_1,\ldots,b_D)$ is the vector of the total number $b_{\ell}$
of links between nodes at distance $d=|q_i-q_j|\in [ d_{\ell-1} ,
d_{\ell}]$ ($d_0=0$). Finally we calculate the entropy of this
ensemble $\Sigma_{\phi(g,\vec{q})}$ and the indicator $\Theta$ from
the definition of Eq. (\ref{Theta}).

\subsection{ The dataset of US airport networks}

\begin{figure}[t]
\includegraphics[width=0.42\textwidth]{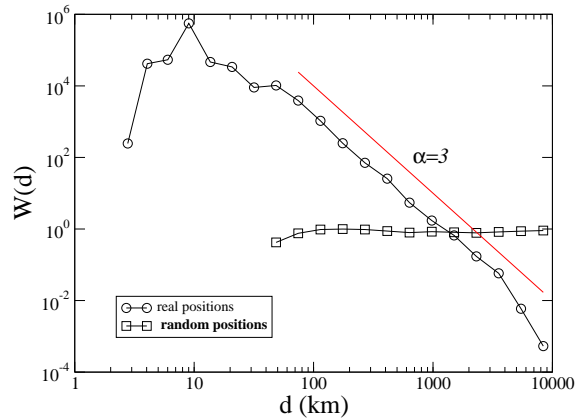}
%\centerline{\psfig{file=Wd.eps,width=7cm}}
\caption{The function $W(q_i,q_j)=W(d)$ in the US airport network, which (see Eq. \ref{community}) encodes the statistical weight
   of a link between airports at distance $d$ (in $km$). For comparison, the same function is shown for the randomized network in which the geographic locations of the airports have been reshuffled. The line, which represents an inverse dependence on the cube of the distance ($\alpha=3$) is drawn as a guide to the eyes. \label{fig3}}
\end{figure}

Here we  apply  the proposed  method to the network
of USA airports studied in \cite{airport2}. We  find that, has it
occur for the internet \cite{Yook}, also  the airport
network is consistent with a power-law dependence of the linking
probability between  two nodes, with their distance.
The network contains $N=675$ airports and $3,253$ connections, each of which is regular
flights between two airports. In this case, with each airport is associated a geographical location $q_i$.
We bin the distances into $D=20$ logarithmically spaced intervals and we
consider as features of our graph  the degree sequence $\vec{k}$ together with
$B(d)$, as discussed above.
We find a high value of $\Theta=1.1 \times 10^3$, showing high significance of space in the structure of airport connections, as expected. In this case, $W(q,q')=W(d(q,q'))$ is a function of the distance only. In Fig. \ref{fig3} we report the shape of the  function $W=W(d)$, depending on
the distance $d$ between any two airports $i$ and $j$, together with the shape of
$WR(d)$ in
the case in which the positions of the airports are randomly
reshuffled.
The function $W(d)$ indicates that the probability  of a
connection decays approximately like a power-law, with deviations for
airports at distances smaller than $100~km$ (flights over such small distances
mainly connect places such as islands or remote areas in Alaska, for which charter flights are the only feasible connection). A log-log fit yields $W(d)\sim d^{-\alpha}$ with $\alpha=3.0\pm 0.2$ for $d>100~km$.

Networks with a linking probability which depends on a power-law of the
distance are of special relevance, both because they occur in real networks (see e.g. \cite{Yook}) and, in abstract terms, for navigability and efficiency \cite{Kleinberg,Boguna,Delos}.

A possible interpretation of the reported statistical regularity is the following. Imagine that flights were designed by a social planner with the aim of optimizing the network for an uniformly distributed population of passengers. This task is similar to that of finding small world networks with optimal navigability. Following the pioneering work of Kleinberg
\cite{Kleinberg}, it has been shown that optimal navigability can be achieved in small-world networks
where long--range links are drawn from a distribution with $\alpha\in [2,3]$ \cite{Delos}.
If we suppose that airports are uniformly distributed across the country and that flying costs have a contribution which increases linearly with distance, then an airline company would face costs
\[
C(R) \propto 2\pi \int_R^{\infty} dr~r^2~W(r) \propto R^{3-\alpha}
\]
to cover distances greater than $R\gg1$. With $\alpha<3$, costs would be dominated by long distance flights.
In a regime of free competition between airlines, $\alpha\ge 3$ is essential in order to maintain a diversified portfolio of flights over short and long distances.
This suggests that $\alpha\approx 3$ corresponds to the optimal compromise between networks with optimal navigability and those which are economically viable in a competitive market of airline companies.

\section{Conclusion}

In conclusion, we propose a method for assessing the relevance of
additional information about the nodes of a networks using the
information that comes from  the topology of
the network itself.
The method makes use of a new quantity $\Theta$, which is not reducible
to any other quantity already introduced in network analysis.
The method can be generalized to directed or weighted networks.
We test and illustrate this method on synthetic as well as real networks, such as the social network of friendship interaction in US schools,
the protein interaction map of {\it Saccharomyces cerevisiae} and the US airport network. As a byproduct, the method also provides additional non--trivial information and highlights hidden statistical regularities.
%We  believe that further applications  will be found useful for the study of other social, biological and technological networks.

\section{Data}

The networks of American schools come from the National Longitudinal Study of
Adolescent Health.
It consists of data from surveys conducted in 1994 in a sample of 84 American high schools
and middle schools by the UNC Carolina Population Center
(http://www.cpc.unc.edu/addhealth).

The protein interaction map that we used is based on the BioGRID database 2.0.20
(http://www.thebiogrid.org). It is described in detail in \cite{Proteins} and is
freely available as
Supplementary Material of \cite{Proteins}.

The airport network was recorded by \cite{airport2} from the 2005 statistics of the
International Air Transport Association (IATA, http://www.iata.org) and is available
at http://cxnets.googlepages.com.

\begin{acknowledgments}
We acknowledge  interesting discussions with Marc Bailly-Bechet,
Marco Cosentino Lagomarsino and Paolo De Los Rios. We
thank Alessandro Vespignani for making the US airports network available to us.
G.B.  acknowledges  support from the  IST STREP GENNETEC contract No.034952.
P.P. and M.M. acknowledge support from EU-STREP project no. 516446 COMPLEXMARKETS.
\end{acknowledgments}

\newpage


\begin{thebibliography}{99}

\bibitem{RMP} A.-L. Barab\'asi and R. Albert, Statistical mechanics of
  complex networks, Rev Mod. Phys,. {\bf 74}, 47-97 (2002).
\bibitem{Doro}
S. N. Dorogovtsev and J. F. F. Mendes, Evolution of Networks: from
biological nets to the Internet and WWW, (Oxford University Press,
Oxford ,2003)
\bibitem{Newman_rev}
M. E. J. Newman, The structure and function of complex networks, SIAM
review {\bf 45}, 157-256 (2003).
\bibitem{Latora}
S. Boccaletti, V. Latora, Y. Moreno, M. Chavez, D.U. Hwang, Complex
networks: Structure and Dynamics. Phys. Rep. {\bf 424}, 175-308
(2006).
\bibitem{Caldarelli}
G. Caldarelli, Scale-free networks, (Oxford University Press,
Oxford,2007).
\bibitem{Monod}
J. Monod, Chance and Necessity: An Essay on the Natural Philosophy of
Modern Biology (WilliamCollins Sons \& Co. Glasgow, 1972).
\bibitem{loopsGBMM}
G. Bianconi and Matteo Marsili, Loops of any size and Hamilton cycles in random
scale-free networks, JSTAT P06005 (2005).
\bibitem{cliquesGBMM}
G. Bianconi and Matteo Marsili, Emergence of large cliques in random scale-free
networks, Europhys. Lett. {\bf 74}, 740 (2006).
\bibitem{Reichardt}
J. Reichardt and S. Bornholdt, Statistical mechanics of community
detection, Phys. Rev. E {\bf 74}, 016110 (2006).
\bibitem{Leone}
M. Leone and A. Pagnani, Predicting protein functions with message
passing algorithms, Bioinformatics {\bf 21}, 239 (2005).
\bibitem{Fortunato_rev}S. Fortunato and C. Castellano,
Community structure in graphs, Springer's Encyclopedia of Complexity
System Science (2008).
\bibitem{Danon}L. Danon, A. D\'iaz-Guilera, J. Dutch and
  A. Arenas,Comparing community structure identification,
JSTAT P09008 (2005).

\bibitem{GN}
M. Girvan and M. E. J. Newman,   Community structure in
social and biological networks. PNAS {\bf 99}, 814 (2002).
\bibitem{Modularity}
M. E. J. Newman and M. Girvan, Finding and evaluating community
structure in networks Phys. Rev. E {\bf 69}, 026113 (2004).
\bibitem{spectral}
M. E. J. Newman, Detecting community structure in networks,
Eur. Phys. J. B {\bf 38}, 321 (2004).

\bibitem{Vicsek}
G. Palla, I. Derenyi, I. Farkas and T. Vicsek, Uncoverisng community
structure of complex networks in nature and society, Nature {\bf 435},
814 (2005).
\bibitem{Newman_hierarchical}
A. Clauset, C. Moore and M. E. J. Newman, Nature {\bf 453}, 98 (2008).
\bibitem{Newman}M. E. J. Newman and E. Leicht, Mixture models and
  exploratory analysis in networks, PNAS {\bf 104}, 9964 (2007).
\bibitem{Arenas}A. Arenas, A. D\'iaz-Guilera and C. J.
  P\'erez-Vicente, Synchronization reveals topological scales in
  complex networks, Phys. Rev. Lett. {\bf 96}, 114102 (2006).
\bibitem{Fortunato}S. Fortunato and M. Barth\'elemy, Resolution limit
  in community detection, PNAS {\bf 104}, 36 (2007).

\bibitem{entropy}
G. Bianconi, The entropy of randomized network ensembles,
Europhys. Lett. {\bf 81}, 28005 (2008).

\bibitem{entropy2}
G. Bianconi, The entropy of network ensembles, Phys. Rev. E {\bf 79},
036114 (2009).
\bibitem{Proteins}
S. Maslov and I. Ispolatov, Propagation of large concentration changes
in revesible protein-binding networks, PNAS {\bf 104}, 13655 (2007).

\bibitem{Moody01}
J. Moody, Race, school integration, and friendship segregation in America, Am. J.
Sociology {\bf 107}(3), 679--716  (2001).
\bibitem{Kertesz}
M. C. Gonzlez, H.J. Herrmann, J. Kertsz, and T. Vicsek,
Physica A {\bf 379}, 307 (2007).
\bibitem{Pin}
S. Currarini, M. O. Jackson and P. Pin, An Economic Model of
Friendship: Homophily, Minorities and Segregation, forthcoming in Econometrica (2008).
\bibitem{Coleman58}
{J. Coleman, Relational analysis: the study of social organizations with survey methods, {\it Human Organization} 17, 28--36 (1958).}
\bibitem{Wright22}
{S. Wright, Coefficients of inbreeding and relationship, {\it American Naturalist} 56, 330--338 (1922).}

\bibitem{Yook}
S. Yook, H. Jeong, A.-L. Barab\'asi, Modeling the Internets large-scale topology, PNAS {\bf 99}, 13382 (2002).

\bibitem{Kleinberg}
J. M. Kleinberg, Navigation in a small world, Nature {\bf 406}, 845 (2000).

\bibitem{Boguna}
M. Bogu\~na, D. Krioukov and K. C. Claffy,Navigability of Complex Networks, Nature Physics {\bf 5}, 74 (2009).

\bibitem{Delos} C. Caretta Cartozo and P. De Los Rios, Extended navigability of small world networks: exact results and new insights, e-print arXiv/0901.4710 (2009).

\bibitem{airport2}
V. Colizza, R. Pastor-Satorras and A. Vespignani, Reaction-diffusion processes and metapopulation models in heterogeneous networks, Nature Physics {\bf 3}, 276-282 (2007).
  
\end{thebibliography}
\end{document}